\pdfoutput=1
\documentclass[12pt,preprint]{aastex}
\usepackage[margin=0.75in]{geometry}
\usepackage{tablefootnote}
\usepackage{graphicx}
\usepackage{natbib}
\usepackage{aas_macros}
\usepackage{subfigure}
\usepackage{float}
\usepackage{color}
\usepackage{footnote}

\def\msun{{\rm\,M_\odot}}

\def\msun{{\rm\,M_\odot}} 
\def\lsun{{\rm\,L_\odot}}

\newcommand{\kms}{\, {\rm km\, s}^{-1}}

\def\h2{${\rm\,H_2}$}

\makeatletter 

\def\kms{{\rm\,km/s}}
\def\msun{{\rm\,M_\odot}}

\def\lsun{{\rm\,L_\odot}}

\def\vol#1  {{{#1}{\rm,}\ }}

\newcount\refno
\newcount\rfno
\def\eq{$^{\the\refno\ }$\advance\refno by 1}
\def\ad{\advance\rfno by 1}

\def\clock{\count0=\time \divide\count0 by 60
     \count1=\count0 \multiply\count1 by -60 \advance\count1 by \time
     \number\count0:\ifnum\count1<10{0\number\count1}\else\number\count1\fi}

\def\myputfigure#1#2#3#4#5%
{\hskip0.03\textwidth\vskip#5pt%\hfill
\makebox[0pt]{\hskip#2in
\includegraphics[width=#3\textwidth]{#1}}\vskip#4pt\hfill}

\newcount\refno
\newcount\rfno
\def\eq{$^{\the\refno\ }$\advance\refno by 1}
\def\ad{\advance\rfno by 1}

\definecolor{burntorange}{rgb}{1,0.4,0.2}

\usepackage{color}

\begin{document}

\title{A New Model for Dark Matter Halos Hosting Quasars}

\author{Renyue Cen\altaffilmark{1} and Mohammadtaher Safarzadeh\altaffilmark{2}}

\footnotetext[1]{Princeton University Observatory, Princeton, NJ 08544;
 cen@astro.princeton.edu}

\footnotetext[2]{Johns Hopkins University, Department of Physics and Astronomy, Baltimore, MD 21218, USA}

\begin{abstract} 

A new model for quasar-hosting dark matter halos, meeting two physical conditions, is put forth. 
First, significant interactions are taken into consideration to trigger quasar activities.
Second, satellites in very massive halos at low redshift are removed from consideration, 
due to their deficiency of cold gas.
We analyze the {\em Millennium Simulation} to find halos that meet these two conditions and 
simultaneously match two-point auto-correlation functions of quasars and 
cross-correlation functions between quasars and galaxies at $z=0.5-3.2$.
%The found halos have some distinct properties worth noting.
The masses of found quasar hosts decrease with decreasing redshift, 
with the mass thresholds being $[(2-5)\times 10^{12}, (2-5)\times 10^{11}, (1-3)\times 10^{11}]\msun$
for median luminosities of $\sim[10^{46}, 10^{46}, 10^{45}]$erg/s at $z=(3.2, 1.4, 0.53)$, 
respectively,
an order of magnitude lower than those inferred based on halo occupation distribution modeling.
In this model quasar hosts are primarily massive central halos at $z\ge 2-3$ but increasingly 
dominated by lower mass satellite halos experiencing major interactions towards lower redshift.
But below $z=1$ satellite halos in groups more massive than $\sim 2\times 10^{13}\msun$ do not host quasars. 
Whether for central or satellite halos, imposing the condition of significant interactions 
substantially boosts the clustering strength compared to the total population with the same mass cut.
The inferred lifetimes of quasars at $z=0.5-3.2$ of $3-30$Myr are in agreement with observations.
Quasars at $z\sim 2$ would be hosted by halos of mass $\sim 5\times 10^{11}\msun$ in this model,
compared to $\sim 3\times 10^{12}\msun$ previously thought, which would help reconcile with
the observed, otherwise puzzling high covering fractions for Lyman limit systems around quasars.

\end{abstract}

\section{Introduction}

Masses of dark matter halos hosting quasars are not directly measured.
They are inferred by indirect methods, such as via their clustering properties 
(i.e., auto-correlation function, ACF, or cross-correlation function, CCF).
Using ACF or CCF can yield solutions on the (lower) threshold halo masses.
The solution on halo mass based on such a method is not unique, to be illustrated by a simple example.
Let us suppose a sample composed of halos of large mass $M$ and an equal number of 
small halos of mass $m$, coming in tight pairs of $M$ and $m$ with 
a separation much small than the scale for the correlation function of interest.
For such a sample, the ACF of halos of mass $M$ 
is essentially identical to that of $m$ or cross correlation between $M$ and $m$.
Although dark matter halos in the standard hierarchical cold dark matter model are less simple,
the feature that small mass halos tend to cluster around massive halos is generic.
This example suggests that alternative solutions of dark matter halos hosting quasars exist. 
It would then be of interest to find models that are based on 
our understanding of the thermal dynamic evolution of gas in halos and other physical considerations,
which is the purpose of this {\it Letter}.
%rather than mathematical selections of dark matter halos. %without reasonable physical backing.
%we use dark matter halos from the {\em Millennium Simulation} \citep[][]{2005dSpringel} 
%to perform an analysis.
%We find that our new models wherein halos have masses much lower (a factor of a few or more) than those in conventional threshold mass based models
%match observed quasar-quasar ACF and quasar-galaxy CCF well.
%The quasar-hosting halos in our model are expected to be cold gas rich at the redshift range examined, $z=0.5$ to $z=3.2$.

%===================
\section{Simulations and Analysis Method}\label{sec: sims}
%===================

We utilize the {\em Millennium Simulation} \citep[][]{2005dSpringel} to perform the analysis, 
whose properties meet our requirements,
including a large box of $500h^{-1}$Mpc,
a mass resolution with dark matter particles of mass $8.6 \times 10^8 h^{-1}\msun$,
and a spatial resolution of 5 $h^{-1}\,{\rm kpc}$ comoving.
%The mass and spatial resolutions are adequate for capturing halos of masses greater than $10^{11}\msun$,
%which are resolved by at least about 100 particles and 40 spatial resolution elements for the virial diameter.
Halos are found using a {\em friends-of-friends} (FOF) algorithm.
Satellite halos %orbiting within each virialized halo 
are separated out using the SUBFIND algorithm \citep[][]{2001dSpringel}. 
The adopted $\Lambda{\rm CDM}$ cosmology parameters are
$\Omega_{\rm m}=0.25$, $\Omega_{\rm b}=0.045$, $\Omega_{\Lambda}=0.75$,
$\sigma_8=0.9$ and $n=1$, and $H_0=100{h\,\rm km} \,{\rm s}^{-1}\,{\rm Mpc}^{-1}$ with $h=0.73$. 
%We do not expect that our results strongly depend on the choice of cosmological parameters
%within reasonable ranges, such as those from \citet{2011Komatsu}.

Given the periodic box we compute the 2-point auto-correlation function (ACF) ${\rm \xi(r_p,\pi)}$
of a halo sample by
\begin{equation}
\label{eq:acf}
{\rm \xi(r_p,\pi) =  \frac{DD}{RR}-1},
\end{equation}
\noindent
where ${\rm r_p}$ and $\pi$ is the pair separation in the sky plane and along the line of sight, respectively,
${\rm DD}$ and ${\rm RR}$ are the normalized numbers of quasar-qausar and random-random pairs in each bin 
${\rm (r_p - \frac{1}{2} \Delta r_p \rightarrow r_p + \frac{1}{2} \Delta r_p, 
 \pi - \frac{1}{2} \Delta \pi \rightarrow \pi + \frac{1}{2} \Delta \pi)}$. 
The cross-correlation function (CCF) is similarly computed:
\begin{equation}
\label{eq:ccf}
{\rm \xi(r_p,\pi) =  \frac{D_1D_2}{R_1R_2} -1},
\end{equation}
\noindent
where ${\rm D_1}$ and ${\rm D_2}$ correspond to
galaxies and quasars. ${\rm R_1}$ and ${\rm R_2}$ 
correspond to randomly distributed galaxies and quasars that are computed analytically.

The projected 2-point correlation function ${\rm w_p(r_p)}$ is:
\citep[][]{1983bDavis} 
\begin{equation}
\label{eq:projected_2pcf}
{\rm w_p(r_p) = 2\int_0^\infty d\pi\ \xi_s(r_p,\pi)}\ .
\end{equation}
\noindent
In practice, the integration is up to $\pi_{\rm max}$.
We use $\pi_{\rm max}=(100,80,70)h^{-1}$Mpc comoving at $z=(3.2,1.4,0.5)$, respectively, as in observations.

\section{A New Model for QSO-Hosting Dark Matter Halos at $z=0.5-3.2$}

\begin{figure}[!h]
\centering
\vskip -0.0cm
\resizebox{4.5in}{!}{\includegraphics[angle=0]{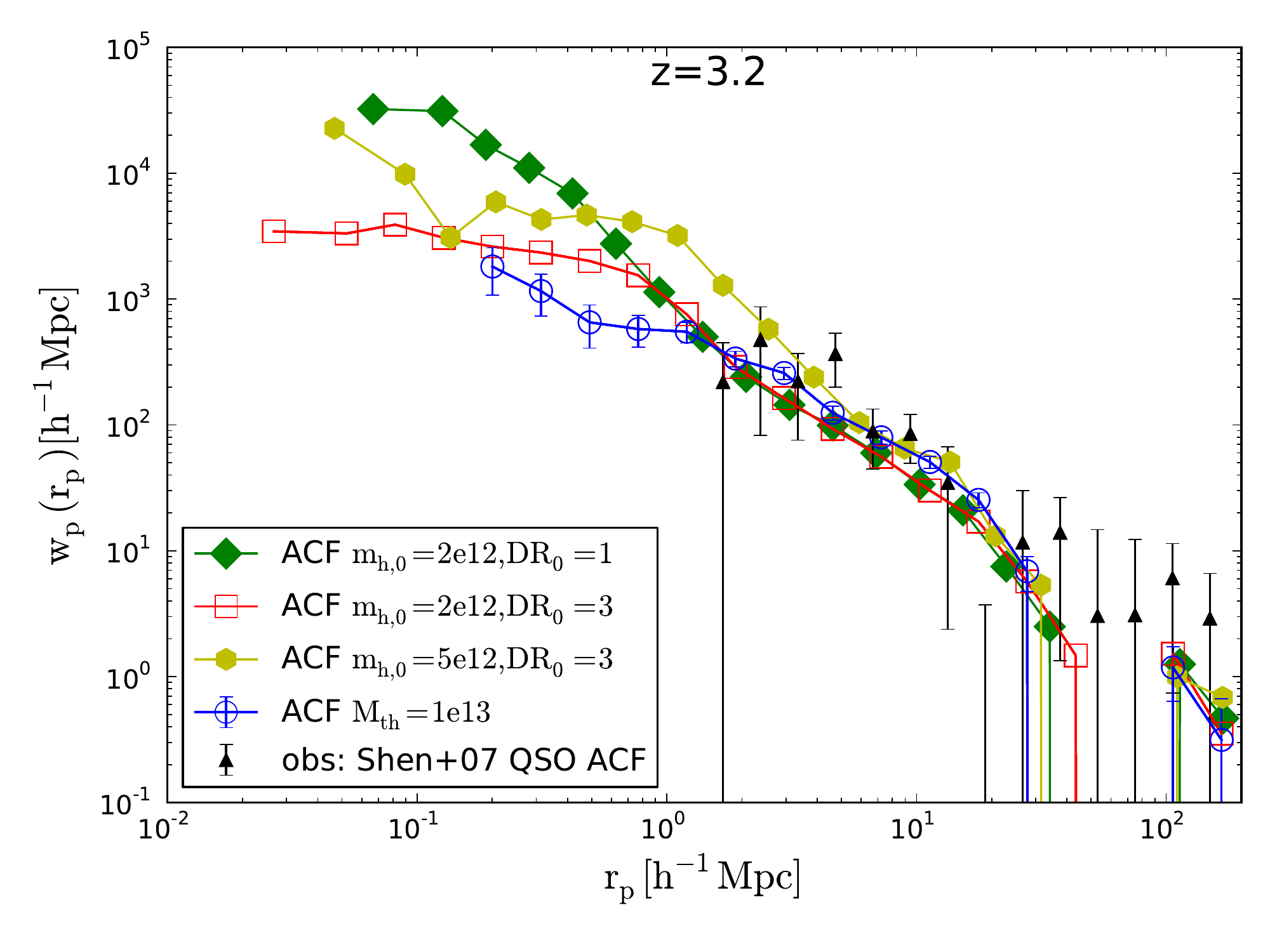}}
\vskip -0.5cm
\caption{%\footnotesize %\scriptsize
shows the ACF of quasar hosts at $z=3.2$ for 
two cases of ${\rm m_{h,0}}=(2\times 10^{12},5\times 10^{12})\msun$ with ${\rm DR_0=3}$ shown as 
(open red squares, solid yellow hexagons), respectively.
For ${\rm m_{h,0}}=2\times 10^{12}\msun$ one additional case is shown for
${\rm DR_0=1}$ (solid green diamonds).
For comparison, a plain threshold mass case with ${\rm M_{th}=10^{13}\msun}$ is shown as open blue circles.
Poisson errorbars are only plotted for blue circles. % for clarity of display.
Black triangles is the observed ACF \citep[][]{2007bShen},
using 4426 spectroscopically identified quasars at $2.9 < z < 5.4$ (median $\bar{z} = 3.2$), 
from the SDSS DR5 \citep[][]{2005aSchneider, 2006aAdelman-McCarthy}. 
}
\label{fig:z3}
\end{figure}

Our physical modeling %of dark matter halos hosting quasars 
is motivated by insights on cosmic gas evolution from cosmological hydrodynamic simulations and observations.
Simulations show four significant trends.
First, cosmological structures collapse to form sheet, filaments and halos, and shock heat the gas 
to progressively higher temperatures with decreasing redshift \citep[e.g.,][]{1999Cen}.  
%resulting in the present-day universe that is dominated by the warm-hot intergalactic medium
%which is observationally verified \citep[][]{2000Tripp,2005Danforth}. 
Second, overdense regions where larger halos are preferentially located
begin to be heated earlier and have higher temperatures than lower density regions at any given time,
causing specific star formation rates of larger galaxies to fall
below the general dimming trend at higher redshift than less massive galaxies
and galaxies with high sSFR to gradually shift to lower density environments at lower redshift.
This physical process of {\it differential} gravitational heating with respect to redshift 
is able to explain the apparent cosmic downsizing phenomenon \citep[e.g.,][]{1996Cowie}, 
the cosmic star formation history \citep[e.g.,][]{2006HopkinsA}, 
and galaxy color migration \citep[][]{2011bCen, 2014Cen}. 
Third, quasars appear to occur in congested environments, as evidenced by high bias inferred based on their strong clustering,
with the apparent merger fraction of bright QSOs ($L>10^{46}$erg/s) approaching unity \citep[e.g.,][]{2014Hickox}.
Finally, a quasar host galaxy presumably channels a significant amount of gas into its central black hole,
which we interpret as the galaxy being rich in cold gas.
This requirement would exclude satellite halos of 
high mass halos at lower redshift when the latter become hot gas dominated \citep[e.g.,][]{2011Feldmann, 2014Cen}. 
These physical considerations provide the basis for the construction of the new model detailed below in steps.
First, for $z>1$.

\begin{figure}[!h]
\centering
\vskip -0.0cm
\resizebox{4.5in}{!}{\includegraphics[angle=0]{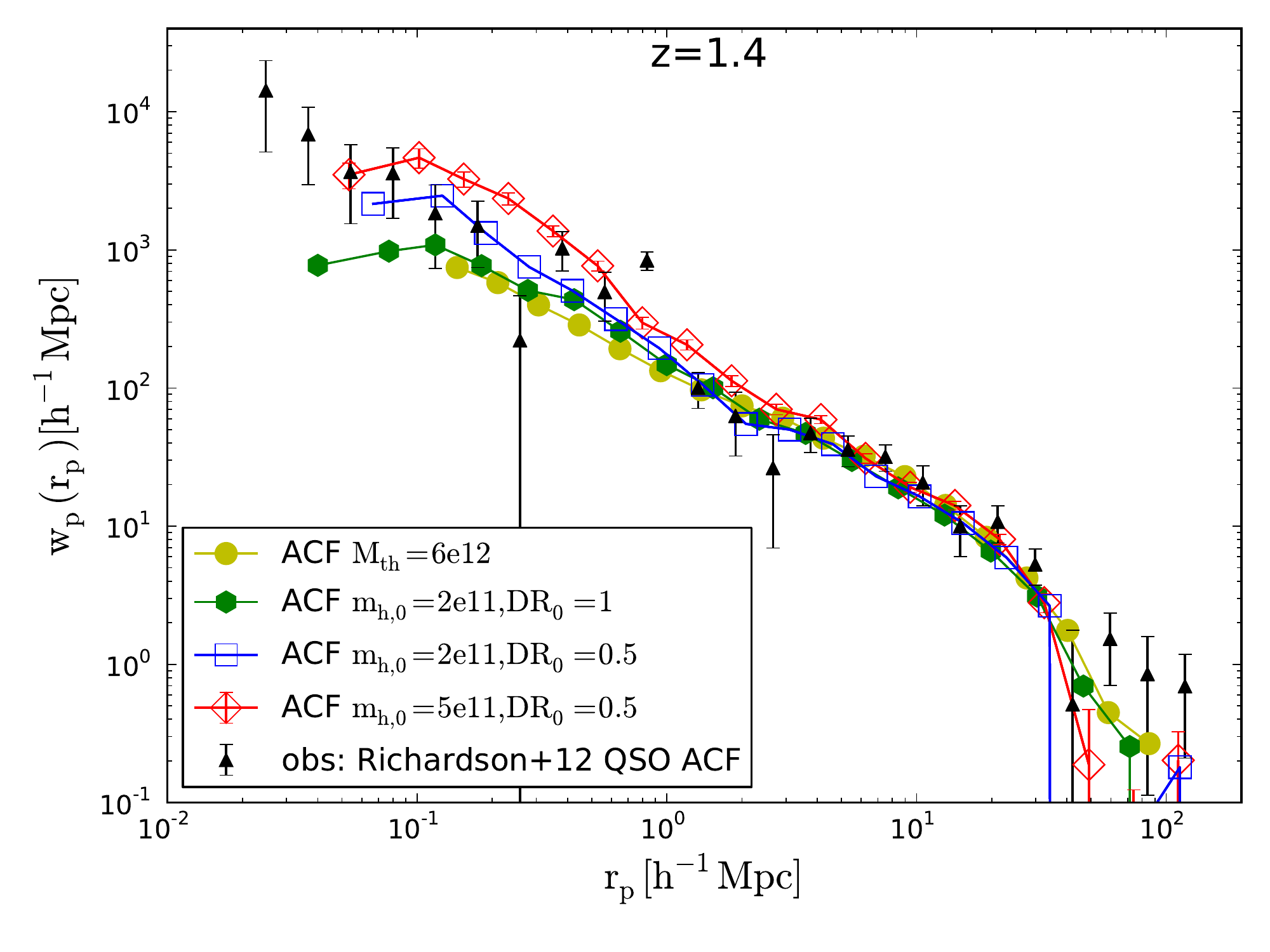}}
\vskip -0.5cm
\caption{%\footnotesize %\scriptsize
shows ACF of quasar hosts at $z=1.4$ for 
three cases: ${\rm (m_{h,0},DR_0)=(2\times 10^{11}\msun, 0.5)}$ (open blue squares),
${(5\times 10^{11}\msun, 0.5)}$ (open red diamonds)
and ${(2\times 10^{11}\msun, 1.0)}$ (solid green hexagons).
For comparison, a plain threshold mass case with ${\rm M_{th}=6\times 10^{12}\msun}$ is shown as solid yellow circles.
Poisson errorbars are only plotted for red diamonds. %for clarity of display.
Black triangles are the observed ACF quasars \citep[][]{2012aRichardson}, 
using a sample of 47,699 quasars with a median redshift of $\bar{z}=1.4$,
drawn from the DR7 spectroscopic quasar catalog \citep[][]{2010aSchneider,2011aShen} for 
large scales and 386 quasars for small scales (${\rm<1}$ Mpc/h) from \citep[][]{2006Hennawi}.
%containing 105,783 spectroscopically confirmed quasars with $M_{i} < -22.0$ over an area of $9380 \; \mathrm{deg^2}$ and a redshift range of $0.065<z<5.46$. 
}
\label{fig:z14}
\end{figure}

\noindent (1) All - central and satellites - halos with virial mass ${\rm >m_{h,0}}$ constitute the baseline sample, denoted as SA.

\noindent (2) Each halo in SA is then selected with the following probability, ${\rm PDF(DR)}$, computed as follows.
For a halo X of mass ${\rm m_h}$, we make a neighbor list of all neighbor halos with mass ${\rm \ge m_h/2}$. %that are within $10$Mpc/h of halo X. 
For each neighbor halo on the neighbor list, we compute ${\rm DR_n=d_n/r_{v,n}}$,
where ${\rm d_n}$ is the distance from X to, and ${\rm r_v}$ is the virial radius of, the neighbor in question.
We then find the minimum of all ${\rm DR_n}$'s, calling it DR for halo X.
${\rm PDF(DR)}$ is defined as 
\begin{equation}
\label{eq:pdf}
{\rm PDF(DR) = 1 \quad for \quad DR<DR_0; \quad PDF(DR)=(DR_0/DR)^3 \quad for \quad DR \ge DR_0}.
\end{equation}
\noindent
Our choice of the specific PDF is somewhat arbitrary but serves to reflect our assertion 
that the probability of dark matter halos hosting quasars decreases if the degree of interactions decreases, when ${\rm DR>DR_0}$.
The results remain little changed, for example, had we used a steeper powerlaw of $4$ instead of $3$.
At $z<1$, when the mean SFR in the universe starts a steep drop \citep[][]{2006HopkinsA},
we impose an additional criterion (3) to account for the gravitational heating. 

\noindent (3) Those halos that are within the virial radius of massive halos ${\rm >M_{h,0}}$ are removed, for $z<1$.

In essence, we model the quasar hosts at $z>1$ with two parameters, ${\rm m_{h,0}}$ and ${\rm DR_0}$
and at $z<1$ with three parameters, ${\rm m_{h,0}}$, ${\rm DR_0}$ and ${\rm M_{h,0}}$.

We present results in the order of decreasing redshift.
Figure~\ref{fig:z3} shows ACF of quasar hosts at $z=3.2$ for three cases:
${\rm (m_{h,0}, DR_0)=(2\times 10^{12}\msun,3)}$, 
$(5\times 10^{12}\msun,3)$ and $(2\times 10^{12}\msun,1)$. 
Based on halo occupation distribution (HOD) modeling, \citet[][]{2012aRichardson} infer median mass of quasar host halos 
at $z\sim3.2$ of $M_{\mathrm{cen}} = 14.1^{+5.8}_{-6.9} \times 10^{12} \; h^{-1} \; \mathrm{M_{\sun}}$,
consistent with the threshold mass case with ${\rm M_{th}=10^{13}\msun}$.
All model ACFs fall below the observed data at ${\rm r_p\ge 30}$Mpc/h, due to simulation box size.
The ACF amplitude is seen to increase with increasing ${\rm m_{h,0}}$.
%Imposing ${\rm DR_0}$ results in a steepening of the ACF below ${\rm r_p\sim 2 DR_0}$.
The ACF with a smaller value of ${\rm DR_0}$ steepens at a smaller ${\rm r_p}$ and rises further toward lower ${\rm r_p}$.
This behavior is understandable, since a lower ${\rm DR_0}$ overweighs pairs at smaller separations.
The extant observations do not allow useful constraints on ${\rm DR_0}$ at $z=3.2$.
We see from visual examination that ${\rm m_{h,0}}=(2-5)\times 10^{12}\msun$ provides an excellent 
fit to the observed ACF for ${\rm r_p=2-30h^{-1}}$Mpc.

Figure~\ref{fig:z14} shows ACF of quasar hosts at $z=1.4$ for 
three cases: ${\rm (m_{h,0},DR_0)=(2\times 10^{11}\msun, 0.5)}$,
${(5\times 10^{11}\msun, 0.5)}$
and ${(2\times 10^{11}\msun, 1.0)}$.
The threshold mass case with ${\rm M_{th}=6\times 10^{12}\msun}$
provides a good match to the observational data for ${\rm r_p=1-30h^{-1}}$Mpc, 
consistent with HOD modeling by \citet[][]{2012aRichardson},
who constrain the median mass of the central host halos to be 
$M_{\mathrm{cen}} = 4.1^{+0.3}_{-0.4} \times 10^{12} \; h^{-1} \; \mathrm{M_{\sun}}$. %and $M_{\mathrm{sat}} = 3.6^{+0.8}_{-1.0}\times 10^{14} \; h^{-1} \; \mathrm{M_{\sun}}$, respectively. 
%The ACF amplitude increases with increasing mass threshold ${\rm m_{h,0}}$.
We see that ${\rm m_{h,0}}=(2-5)\times 10^{11}\msun$ provides excellent fits to the observed ACF for
${\rm r_p=1-40h^{-1}}$Mpc. 
The observed ACF extends down to about $20h^{-1}$kpc, which allows us to constrain ${\rm DR_0}$.
We see that, varying ${\rm DR_0}$ from $1.0$ to $0.5$,
the amplitude of the ACF at ${\rm r_p\le 1h^{-1}}$Mpc increases,
with ${\rm DR_0=0.5}$ providing a good match.
%While possible to add complexities to try matching the observed points at small ${\rm r_p}$,
%it is not useful at this point.
%Nonetheless, 
The physical implication is that quasar activities at $z=1.4$ seem to be triggered when
a halo of mass $\ge (2-5)\times 10^{11}\msun$ interact significantly with another halo of comparable mass,
in contrast to the $z=3.2$ quasars that are primarily hosted by central galaxies with no major companions.

\begin{figure}[!h]
\centering
\vskip -0.0cm
\hskip -1.65cm
\resizebox{3.76in}{!}{\includegraphics[angle=0]{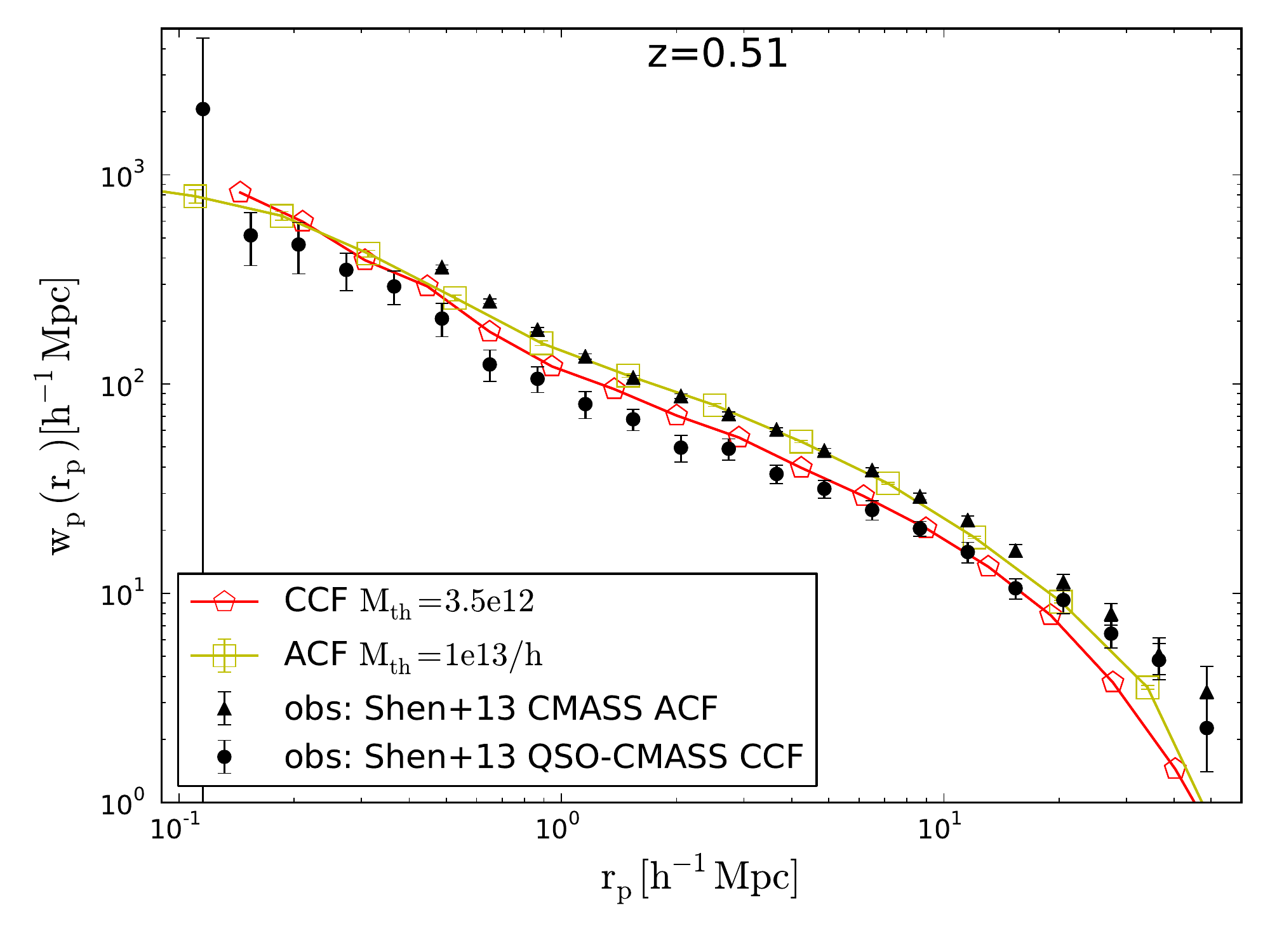}}
\hskip -0.00cm
\resizebox{3.76in}{!}{\includegraphics[angle=0]{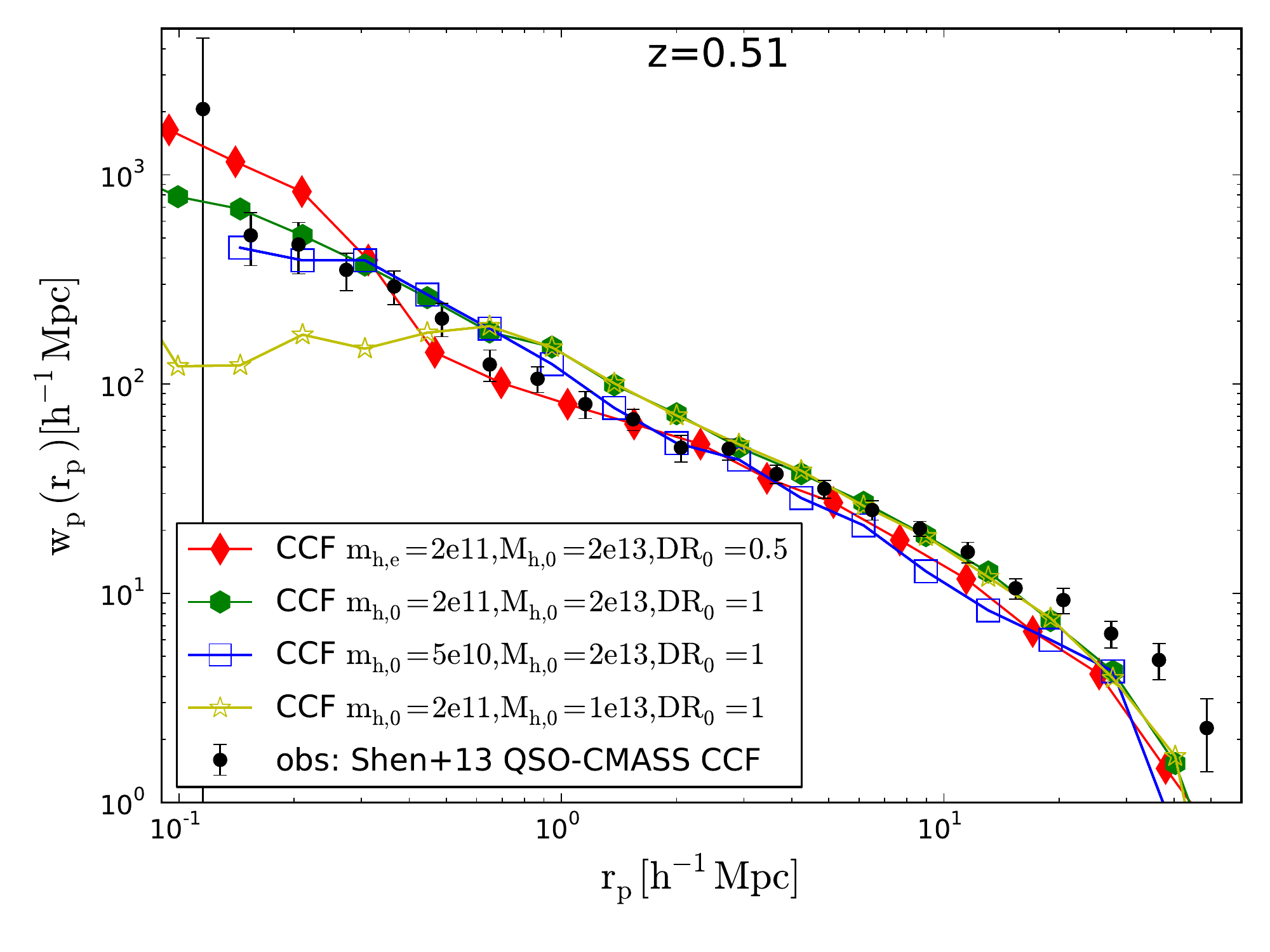}}
\vskip -0.0cm
\caption{%\footnotesize %\scriptsize
{\color{burntorange}\bf Left panel}
shows the ACF of halos of masses above ${\rm M_{th}=1\times 10^{13}h^{-1}\msun}$ (open yellow squares), ``mock CMASS galaxies",
and the CCF between halos of mass above $3.5\times 10^{12}\msun$ and CMASS galaxies (open red pentagons).
Black solid dots and triangles are the observed quasar-CMASS galaxy CCF 
and CMASS galaxy ACF (shown in both left and right panels), respectively, at $z\sim 0.53$ from \citet[][]{2013Shen}.
The CMASS sample of 349,608 galaxies at $z\sim 0.53$ \citep[][]{2011aWhite,2012cAnderson} 
is from the Baryon Oscillation Spectroscopic Survey \citep[][]{2009aSchlegel,2013aDawson}.
%The characteristic halo mass of the CMASS sample has been previously found to be $\sim 10^{13} h^{-1} M_{\odot}$, which is verified here.
The sample of 8198 quasars at $0.3<z<0.9$ ($\langle z\rangle\sim 0.53$) is from the DR7 \citep{2009aAbazajian}
spectroscopic quasar sample from SDSS I/II \citep[][]{2010aSchneider}.
{\color{burntorange}\bf Right panel}
shows the model quasar-CMASS galaxy CCF at $z=0.51$ for 
four cases with ${\rm (m_{h,0}, M_{h,0}, DR_0) =(2\times 10^{11}\msun, 2\times 10^{13}\msun,0.5)}$ (solid red diamonds),
${\rm (2\times 10^{11}\msun, 2\times 10^{13}\msun, 1.0)}$ (solid green hexagons),
${\rm (5\times 10^{10}\msun, 2\times 10^{13}\msun, 1.0)}$ (open blue squares)
and 
${\rm (2\times 10^{11}\msun, 1\times 10^{13}\msun, 1.0)}$ (open yellow stars).
}
\label{fig:z05}
\end{figure}

Finally, Figure~\ref{fig:z05} shows results at $z=0.51$.
The left panel shows the ACF of halos of masses above the threshold $10^{13}h^{-1}\msun$ -  mock CMASS galaxies -
which provides a good match to the observed ACF of CMASS galaxies.
Consistent with previous analysis, we see that the CCF between halos of mass above the threshold $3.5\times 10^{12}\msun$ and mock CMASS galaxies 
match the observed counterpart.
%Having made these verifications, 
The right panel of Figure~\ref{fig:z05} shows 
the mock quasar-CMASS galaxy CCF at $z=0.51$ for four cases with different combinations of ${\rm (m_{h,0}, M_{h,0}, DR_0)}$.
The case with ${\rm (m_{h,0}, M_{h,0}, DR_0) =[(1-3)\times 10^{11}\msun, 2\times 10^{13}\msun,0.5]}$
provides an adequate match to the observation, while
${\rm (m_{h,0}, M_{h,0}, DR_0) =(5\times 10^{10}\msun, 2\times 10^{13}\msun,1.0)}$
appears to underestimate the CCF.
%What is striking is the large difference between 
%${\rm (m_{h,0}, M_{h,0}, DR_0) =(2\times 10^{11}\msun, 2\times 10^{13}\msun,0.5)}$
The case with ${\rm (2\times 10^{11}\msun, 1\times 10^{13}\msun, 1.0)}$
significantly underestimates the observed ACF at ${\rm r_p<0.5h^{-1}}$Mpc.
This indicates that halos of masses greater than ${\rm m_{h,0} =(1-3)\times 10^{11}\msun}$ 
residing in environment of groups of masses $(1-2)\times 10^{13}\msun$ are primarily responsible
for the strong clustering at ${\rm r_p<0.5h^{-1}}$Mpc.
It is interesting to note that the exclusion halo mass of ${\rm M_{h,0}}=2.0\times 10^{13}\msun$, to account for environment heating effects,
is physically self-consistent with the fact that the red CMASS galaxies are red due to the 
same environment effects hence have about the same halo mass (${\rm M_{th}=10^{13}h^{-1}\msun}$).

\section{Predictions and Tests of our Model}

\begin{figure}[!h]
\centering
\vskip -0.0cm
\resizebox{4.5in}{!}{\includegraphics[angle=0]{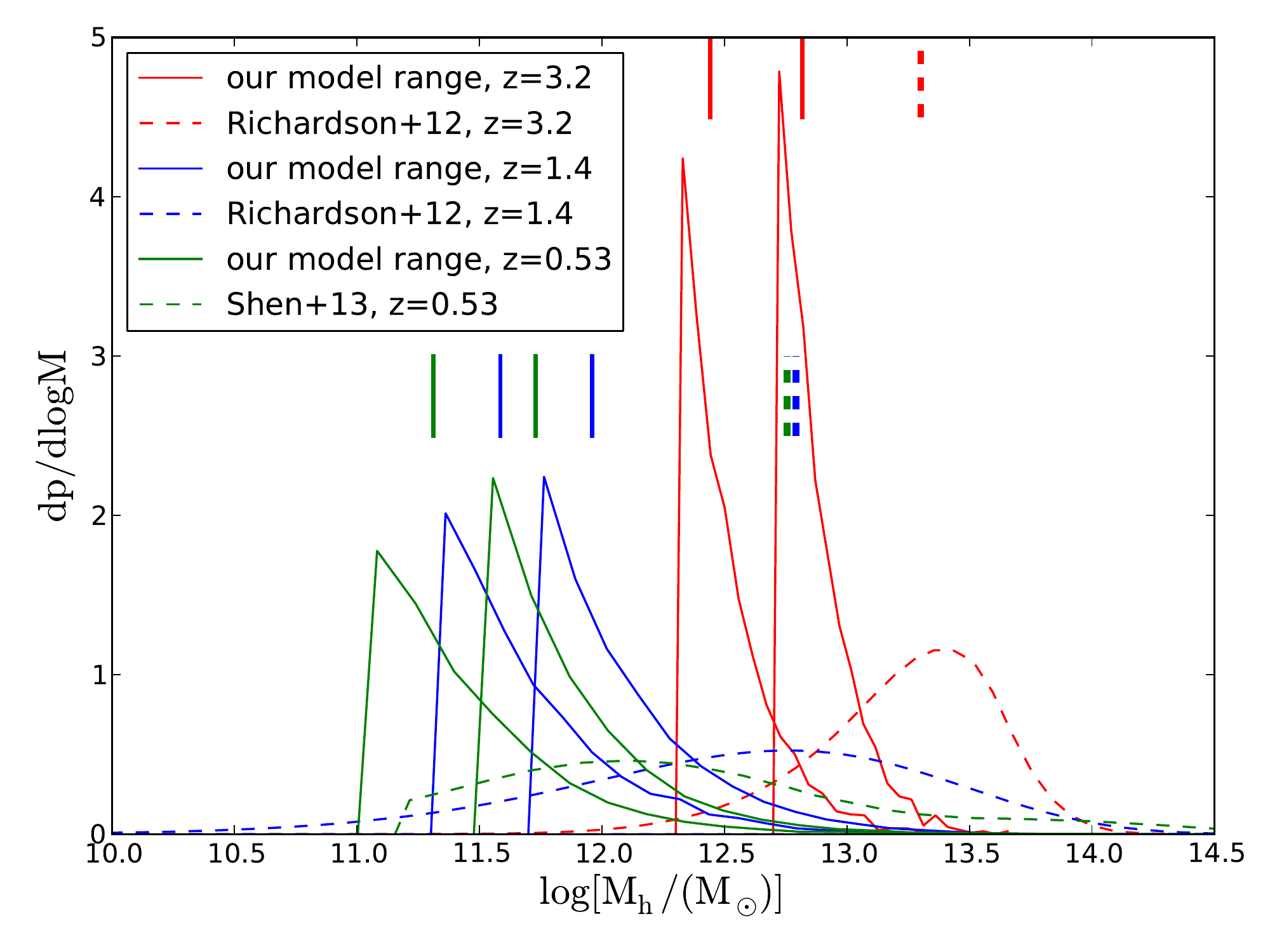}}
\vskip -0.5cm
\caption{%\footnotesize %\scriptsize
shows the QSO-hosting halo mass distributions at $z=3.2$ (solid red curves),
$z=1.4$ (solid blue curves)
and $z=0.53$ (solid green curves).
We show two bracketing (approximately $\pm 1\sigma$ for the computed correlation functions) models at each redshift.
The corresponding distributions based on HOD modeling are shown in dashed curves.
The short vertical bars with matching colors and line types indicate the median halo masses of their respective distributions.
}
\label{fig:mass}
\end{figure}

We have demonstrated that our physically based model can account for the observed clustering of quasars at $z=3.2, 1.4, 0.53$.
Figure~\ref{fig:mass} contrasts the sharp differences between our model and the conventional HOD based modeling;
the halo masses in our model are an order of magnitude lower than those inferred from HOD modeling.
%We also predict that the average masses of quasar hosts 
%decrease with decreasing redshift. %diametrically opposed to previous models.
Our model gives quasar-hosting halo mass threshold of
$[(2-5)\times 10^{12}, (2-5)\times 10^{11}, (1-3)\times 10^{11})]\msun$ at $z=(3.2, 1.4, 0.53)$, respectively.
compared to median mass of $(14.1^{+5.8}_{-6.9} \times 10^{12}, 4.1^{+0.3}_{-0.4} \times 10^{12},  4.0 \times 10^{12})h^{-1}\msun$ 
based on HOD modeling \citep[][]{2012aRichardson, 2013Shen}.
Although we have not made fitting for quasars at redshift higher than $z=3.2$,
we anticipate that the quasars at higher redshifts 
that have comparable luminosities as those at $z=3.2$ will primarily be hosted
by central galaxies of mass $\sim (2-5)\times 10^{12}\msun$.
We note that the median luminosity of the observed 
quasars decreases from $\sim 10^{46}$erg/s at $z\ge 1.4$ to $\sim 10^{45}$erg/s at $z=0.53$, 
which reflects the known downsizing scenario and is in accord with the decreasing halo mass with decreasing redshift inferred 
in our model.
Our results and detailed comparions with HOD based modeling are also tabulated in Table 1,
along with inferred quasar duty cycles and lifetimes.

Can we differentiate between these two models?
%The large differences in inferred halos masses potentially provide the most discriminatory tests of our model.
%A most direct test of our model is to obtain dynamically the halo masses of observed quasars,
%rather than inferred from clustering analysis.
%Some recent studies have provided independent hints in favor of our model.
\citet[][]{2012Trainor}
cross correlate 1558 galaxies with spectroscopic redshifts with 15 of the most 
luminous ($\ge 10^{14}\lsun, M_{1450} \sim -30$) quasars at $z \sim 2.7$. 
Even for these hyperluminous quasars (HLQSOs), they infer host halo mass of ${\rm \log (M_h/\msun) = 12.3 \pm 0.5}$, 
which is in very good agreement with our model ($M_h\sim (2-5)\times 10^{12}\msun$)
but much smaller than inferred from HOD modeling.
They also find that, on average, the HLQSOs lie within significant galaxy over-densities, characterized by a velocity dispersion 
$\sigma_v \sim 200\kms$ and a transverse angular scale of $\sim$25'' ($\sim$200 physical kpc), which they argue correspond 
to small groups with ${\rm \log (M_h/\msun) \sim 13}$. 
The rare HLQSOs are apparently not hosted by rare dark matter halos.
This is fully consistent with our suggestion that dark matter halo mass
is not the sole determining factor of quasar luminosities
and that interactions may be instrumental to triggering quasar activities.

Another, independent method to infer halo masses of quasar hosts is to measure their cold gas content.
\citet[][]{2013Prochaska} detect about $60\%-70\%$ covering fraction of Lyman limit systems 
within the virial radius of $z\sim 2$ quasars, using the binary quasar sample \citep[][]{2006Hennawi}.
This has created significant tension: hydrodynamic simulations of the cold dark matter model yield less than
20\% covering fraction for halos of mass $\sim 3\times 10^{12}\msun$ \citep[][]{2014FG};
halos of still higher mass have still lower covering fractions.
On the other hand, the simulations show a $\sim 60\%$ covering fraction if the mass of quasar-hosting halos is $\sim 3\times 10^{11}\msun$.
This indicates that the lower halo masses for quasar hosts
in our model can explain the high content of neutral gas in $z\sim 2$ quasars.

The mean quasar lifetime may be estimated by equating it to 
${\rm t_H \times f_q}$, where ${\rm t_H}$ is the Hubble time at the redshift in question
and ${\rm f_q}$ the duty cycle of quasar hosting halos.
Existing observational constraints provide useful range for $t_{\rm q}$ for quasars at $z\sim 3$.
%Among a host of studies utilizing various observations,
%aside from the constraint of $t_{\rm QSO} \ge 60$Myr at $z\sim 3$ by \citet[][]{2007Cantalupo},
%$t_{\rm QSO}=10-100$Myr appears to be consistent with the general consensus
Lifetimes based on halo abundances from clustering analyses of quasars 
have been given by many authors %(including this study)
\citep[e.g.,][]{2001Martini, 2004Porciani, 2007Shen};
in our case, this is a degenerate derivation.
Thus, it is useful to have a survey of quasar lifetimes based on other, independent methods.
\citet[][]{2003Jakobsen} derive ${\rm t_q>10}$Myr,
\citet[][]{2007Worseck} give ${\rm t_q>25}$Myr,
\citet[][]{2008Goncalves} yield ${\rm t_q=16-33}$Myr,
and 
\citet[][]{2014McQuinn} yield ${\rm t_q=>10}$Myr 
for quasars at $z\approx 2-3$, all based on the method of quasar proximity effect.
\citet[][]{2012Bolton} obtain ${\rm t_q>3}$Myr using line-of-sight thermal proximity effect.
\citet[][]{2013Trainor}, using a novel method of Ly$\alpha$ emitters (LAEs) exhibiting fluorescent emission via the
reprocessing of ionizing radiation from nearby hyperluminous QSOs, 
find ${\rm 1 \le t_q \le 20}$Myr at $z=2.5-2.9$.
We see that all these estimates are consistent with our model. 
As a comparison, the inferred ${\rm t_q^{HOD}\sim 400}$Myr at $z=3.2$ from HOD modeling.
%is equal to about 10 Salpeter times or a factor of $\ge 10^4$ growth in black hole mass per Hubble time,
%which might result in black holes that are too massive, unless these massive hosts start with 
%a small black hole seed ($\ll 10^5\msun$).

Finally, self-consistently reproducing the quasar luminosity functions 
\citep[e.g.,][]{2002Wyithe, 2003Wyithe, 2009Shen, 2013Conroy}
will provide another test, which we defer to a separate study.

\begin{table*}[h]
\caption{Comparing Our Model With HOD Modeling w.r.t. Halo Mass and Quasar Lifetime}
\begin{tabular}{llccccccccc}
\hline
(1)& (2) & (3)& {\bf (4)} &{\bf (5)}& {\bf (6)}& {\bf (7)} & {\vrule}& (8) & (9) & (10) \\
${\rm z_{\rm med}}$& ${\rm \ L_{bol} }$ &${\rm n_{obs}}$&${\bf n_{sim}} $ &
${\bf m_{h,0}}$ &  ${\bf f_q}$ & ${\bf t_q}$ & {\vrule}& ${\rm M^{HOD}_{h} }$ & ${\rm f_q^{HOD}}$ & ${\rm t_q^{HOD}}$ \\
 &${\rm \log (\frac{erg} {s})}$  &${\rm \times 10^{-7}}$&${\bf \times 10^{-4}}$&${\bf \times 10^{12}}$& ${\bf \times 10^{-3}}$&{\bf [Myr]}  & {\vrule}& ${\rm \times 10^{12}}$ &${\rm \times 10^{-3}}$  &[Myr] \\
\hline
 3.2&46.3& ${\rm 2.5}$&${\bf 0.2-0.9}$&${\bf 2-5 }$& ${\bf 3-13}$&{\bf 5-26}&{\vrule}&${\rm 20}$ &  ${\rm 215}$&425\\
1.4&46.1&${\rm 30}$&${\bf 9-44}$& ${\bf 0.2-0.5}$&${\bf 0.6-3}$&{\bf 3-15}&{\vrule}&${\rm 5.8}$  &${\rm 1.8}$&7.5\\
0.53&45.1&${\rm 50}$&${\bf 29-85}$&${\bf 0.1-0.3}$&${\bf 0.6-2}$&{\bf 5-15}&{\vrule}&${\rm 5.7}$   &  ${\rm 1.3}$  &10\\
\hline
\end{tabular}\label{tab:summary}
\vskip 0.3cm
{Column (1) ${\rm z_{\rm med}}$ is the median redshift of the sample that is analyzed.\\
Column (2) ${\rm \ L_{bol}} $ is the median bolometric luminosity of the observed quasar sample obtained using conversions in \citep[][]{2006bRichards,2012aRunnoe}.\\
Column (3) ${\rm n_{obs}}$ is the number density of the observed quasar sample \citep[][]{2012aShen} in ${\rm [Mpc^3 h^{-3}]}$, multiplied by a factor of $2.5$ to account
for the fact that about 60\% of quasars belong to the so-called type II quasars based on low redshift observations \citep[][]{2003Zakamska}, 
which are missed in the quoted observational sample. We note that the percentage of obscured quasars appear to increase with redshift \citep[e.g.,][]{2006Ballantyne,2006Treister}.
Thus, the values of ${\rm t_q}$ may be underestimated.\\
Column (4) ${\rm m_{h,0}}$ is the lower mass threshold dark matter halos hosting quasars in ${\rm [M_\odot]}$.\\
Column (5) ${\rm n_{sim}}$ is the number density of the dark matter halos hosting quasars with mass $\ge {\rm m_{h,0}}$ in ${\rm [Mpc^{-3} h^{3}]}$.\\
Column (6) ${\rm f_q}\equiv n_{\rm obs}/n_{\rm sim}$ is the duty cycle of the quasars in our model.\\
Column (7) ${\rm t_q}$ is the mean quasar lifetime in our model defined as ${\rm t_H \times f_q}$, where ${\rm t_H}$ is the Hubble time at the redshift in question.\\
Column (8) ${\rm M^{HOD}_{h,med}}$ is the derived host halo mass of the observed population of quasars derived from HOD modeling \citep[][]{2012aRichardson,2013Shen} in ${\rm [M_\odot]}$. \\
Column (9) ${\rm f_q^{HOD}}$  is the duty cycle of the observed population of quasars based on HOD modeling, using
the type II quasars-corrected abundance in Column (3). \\
Column (10) ${\rm t_q^{HOD}}$  is the life time of the quasars based on ${\rm f_q^{HOD}}$ in Column (9).
} 
\label{tab:title}
\end{table*}

\section{Conclusions}

We put forth new model for dark matter halos that host quasars.
Our model is substantially different from previous models based on simple lower mass threshold or HOD based lower mass threshold.
Instead, we impose two conditions that are physically based.
First condition is that significant interactions with other halos are a necessary ingredient to trigger quasar activities.
Second, satellite halos within the virial radius of large halos above certain mass 
at low redshift are removed from consideration, since they are hot gas dominated.
We investigate this model utilizing halo catalogs from the {\em Millennium Simulation}.
By requiring simultaneously that halos meet these two conditions and 
match two-point auto-correlation functions of quasars and cross-correlation functions between quasars and galaxies,
we are able to identify quasar hosting halos.
The resulting host halos are distinctly different from other models.
Quasar hosts are less massive, by an order of magnitude, than inferred based on either simple halo mass threshold or HOD models.
%Second,  the typical masses of quasar hosts decrease with decreasing redshift. 
The lower halo mass threshold of quasar hosts are predicted to be 
$[(2-5)\times 10^{12}, (2-5)\times 10^{11}, (1-3)\times 10^{11}]\msun$ at $z=(3.2, 1.4, 0.53)$, respectively,
compared to median mass of $(14.1^{+5.8}_{-6.9} \times 10^{12}, 4.1^{+0.3}_{-0.4} \times 10^{12},  4.0 \times 10^{12})h^{-1}\msun$ 
based on HOD modeling.
It is noted that, for either central or satellite halos, imposing the condition of significant interactions 
substantially boosts the clustering strength compared to the total population with the same mass cut,
which is the reason why our lower mass halos can equally well match observed clustering of quasars.
At $z\ge 2$ the quasar hosts are primarily central galaxies and major interactions at close separations do not appear to be necessary,
whereas at $z<2$ the quasar hosts mostly are satellite galaxies which experience major interactions to trigger the quasar activities.
Below $z=1$ satellite galaxies in groups of mass $\ge 2\times 10^{13}\msun$ do not host quasars due to lack of cold gas.
The mean quasar lifetime is, nearly invariably, expected to be in the range of $3-30$Myr for the redshift range examined $z=0.5-3.2$,
which is in good agreement with observations.
A unique and discriminatory property is that, unlike other previous models, 
the quasar hosting galaxies in this model are expected to be cold gas rich and have much higher
covering fractions for Lyman limit systems, better in line with observed
$60-70\%$ covering fraction within the virial radius of $z\sim 2$ quasars.

%\acknowledgments
\vskip 1cm

We are grateful to Dr. Zheng Zheng for allowing us to his program to ACF and CCF,
to Dr. Jonathan Richardson and Dr. Yue Shen 
for sending us the observed ACF and CCF at $z=1.4$ and $z=3.2$, and at $z=0.53$, respectively.
We also would like to thank Dr. Zheng for useful discussion.
This work is supported in part by grant NASA NNX11AI23G.
The Millennium simulation data bases used in this paper and the web application 
providing online access to them were constructed as part of the activities of the German Astrophysical Virtual Observatory.

%\bibliographystyle{apj}
%\bibliography{astro}

\end{document}